\documentclass[runningheads,citeauthoryear]{apinv46}
\usepackage{epsfig,cite,graphics}
\usepackage[utf8]{inputenc}
\usepackage{xcolor}

\begin{document}

\title{$BV$ photometry of  the cataclysmic variable LY~UMa }
\titlerunning{LY UMa}
\author{L. Dankova\inst{1}, D. Marchev\inst{2}, 
        S. Peneva\inst{1}, R. Zamanov\inst{1}}
\authorrunning{Dankova et al. }
\tocauthor{L. Dankova, D. Marchev, S. Peneva, R. Zamanov} 
\institute{\vskip 0.2cm $^{} \;$ \\
$^1$Institute of Astronomy and National Astronomical Observatory, Bulgarian Academy of Sciences, Tsarigradsko Shose 72, BG-1784 Sofia, Bulgaria \\
\vskip 0.2cm
$^2$Department of Physics and Astronomy, Shumen University "Episkop Konstantin
Preslavski", 115 Universitetska Str., 9700 Shumen, Bulgaria
	\vskip 0.1cm 
	\email{e-mails: speneva@nao-rozhen.org  rzamanov@nao-rozhen.org \vskip 0.1cm }  
	} 
\papertype{Research report, submitted on 07.01.2026;  accepted on 01.08.2026 \\ 
 {\color{green} www.astro.bas.bg/AIJ/issues/n46/index.php}}
	

\maketitle

\begin{abstract}
We performed $B$- and $V$-band photometry of the probable dwarf nova LY~UMa.
We find that it has apparent  magnitude 
    $V = 15.26 \pm 0.03$ on 2024-04-14,
    $V = 15.43 \pm 0.02$ on 2025-04-28,
and $V = 15.57 \pm 0.02$ on 2025-06-02. 
Using the GAIA distance $d=310$~pc, we find that 
the lower values  correspond to absolute $V$ band magnitude $M_V \approx 7.94-8.10$ and
indicate that the mass donor of LY~UMa is a K7-K8V star with $M_V \approx 8.0-8.5$. 
Our $B$-band measurements show slightly bluer $B-V$ index of LY UMa 
than that of its donor, consistent with its nature as a cataclysmic variable. \\
Close to LY~UMa is the Active Galaxy Nucleus Mrk 153, for which we estimate $V=14.81  \pm 0.03$. 
\end{abstract}
\vskip 0.2cm 

\section{Introduction}

Cataclysmic variable stars (CVs)  are so named 
because they irregularly increase in brightness (sometimes by factor of 10 - 100 or even larger), then drop back down to a quiescent state. 
CVs consist of two components: 
a white dwarf and a mass donor star. 
They  are divided into several sub-types -- nova-like variables, dwarf novae, magnetic, 
AM~CVn binaries, classical novae,  recurrent novae, etc.  (Warner 1995).  

CVs are a result of the evolution 
of pairs of main-sequence stars. 
The white dwarf components of  CVs
were once the cores of giant stars much larger than CVs are now. 
The standard theory for the evolution of binary stars
requires that the system passes through a phase during which both stars orbit
within a common envelope (a result of the expansion of the giant star). 
As the stars orbit, they transfer angular momentum to the envelope. 
The envelope is ejected and leaves a binary star with short orbital period (a few hours),
e.g.  Marsh (2000), Solheim (2010) and references therein. 

LY~UMa (CW~1045+525) is a variable star in the constellation 
Ursa Major. It is an emission-line binary 
with a strong late-K contribution and no reported outbursts. 
The orbital period is $P_{orb} = 0.2712788$~days, which is 6.51~hours 
(Tappert et al. 2001, Thorstensen et al. 2017). 


Here we report $B$- and $V$-band photometry of LY~UMa.  Close to LY~UMa
is located Mrk~153 and its $V$-band magnitude is also estimated.

 \begin{figure}    
   \vspace{11.7cm}     
   \includegraphics{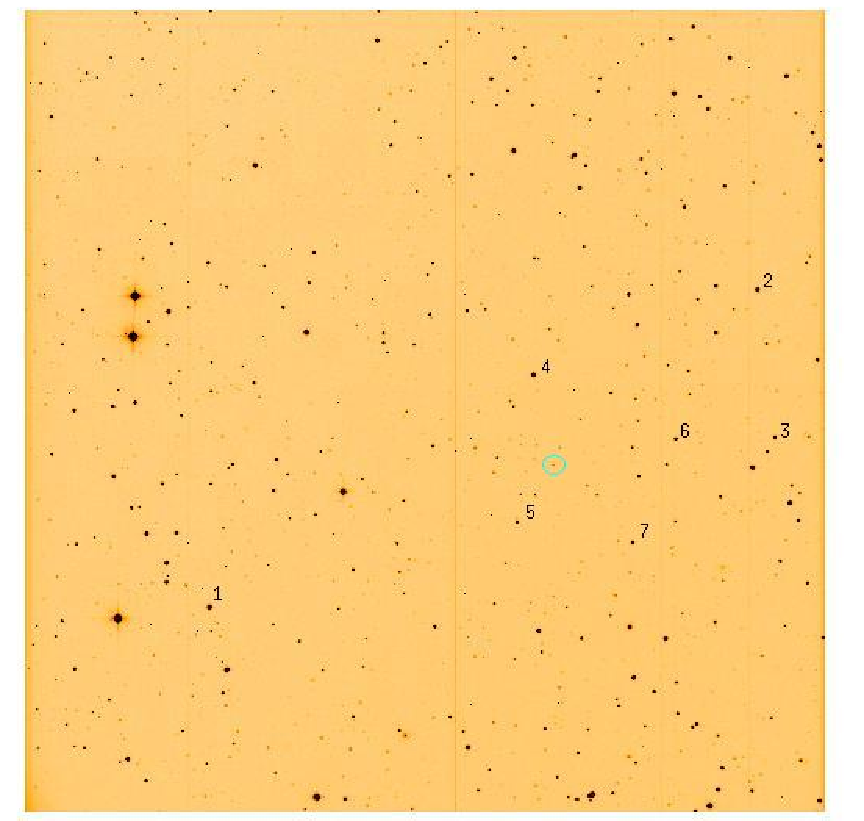}  
   \caption[]{$V$ band image of the field around LY~UMa, obtained with the 50/70~cm Schmidt 
    telescope of NAO Rozhen on 2025-04-28T21:45, exposure time  60 sec. 
    The field of view is 61x61 arcmin. 
    LY~UMa is marked with a circle. The comparison stars are marked with 
    numbers from 1 to 7. Their coordinates and magnitudes 
    can be found in Table~\ref{t.cs}. 
     }
  \label{f.1}      
  \end{figure}        

\section{Observations and data processing} 
\label{s.obs}

The observations were secured with the 2.0~m and the 50/70~cm Schmidt telescopes of
the Rozhen National Astronomical Observatory, Bulgaria 
(Golev et al.  1982; Tsvetkov et al. 1987) and 
with the new 60~cm Ritchey-Chretien telescope of the 
Shumen University Astronomical Center (Marchev et al. 2025). 
The telescopes are equipped with CCD cameras: 
Andor iKON-L (2.0~m telescope),  FLI~PL~16803 (50/70~cm Schmidt telescope), 
and SBIG Aluma4040 (60~cm Shumen telescope).

LY~UMa  has coordinates J2000 10:48:18.0 +52:18:29. 
The object is located in a direction (Galactic  coordinates ep=J2000: 156.88544 +55.9249) 
with no  interstellar dust 
and several quasars and galaxies can be seen in the field. 
The brightest among them is Mrk~153.

 \begin{figure}    
   \vspace{7.0cm}     
   \includegraphics{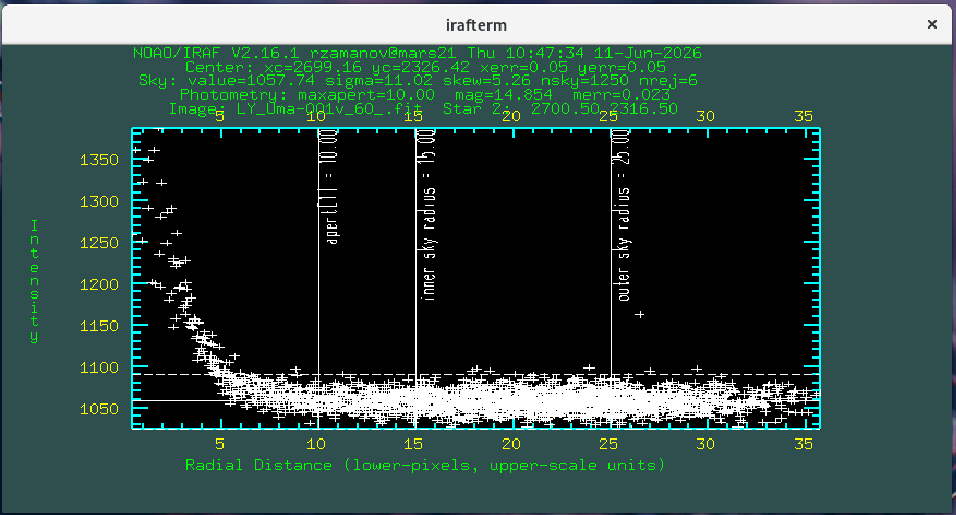}  
   \caption[]{Example of photometry with $qphot$ in IRAF. } 
  \label{f.2}      
  \end{figure}        

The field of view of the 50/70~cm Schmidt telescope is 61x61 arcmin, with CCD frame 
4096x4096 pixels; this gives 0.9 arcsec/pixel. 
For the data processing we used IRAF (Tody 1993).  
Typical level of the bias of our CCD frames is  $988 \pm 8$ counts. 
Averaging  8 bias images,  we achieve  $988 \pm 4$ counts. 
Here we give some of the critical parameters used in $qphot$ of IRAF: \\
$ \; $ \\
{\small
\begin{tabular}{l l l l }
$\; $ & cbox    =		      30.  &  & The centering box width in pixels \\
$\; $ & annulus =		      15.  &  & The inner radius of sky annulus in pixels \\
$\; $ & dannulus=		      10.  &  & The width of the sky annulus in pixels  \\
$\; $ & aperture=		       10  &  & The list of photometry apertures  \\
$\; $ & (zmag   =		      25.) &  & The zero point of the magnitude scale  \\
$\; $ & (interac=		      yes) &  & Interactive mode ?  \\
$\; $ & (radplot=		      yes) &  & Plot the radial profiles in interactive mode ?  \\
\end{tabular}\\
}

Since we use a few comparison stars in the field, the 
zero-point of the magnitude scale was set to 25 by default. 
The apertures used are 7 to 10 pixels (corresponding to 6-9 arcsec), 
the inner radius of sky annulus
is usually set to 1.5 apertures, the width of the sky annulus is equal to the aperture,
the centering box width is 3-4 apertures. An example of the photometry with IRAF is given 
in Fig.~\ref{f.2}. The figure shows that at exposure time of 60 sec, 
the sky  brightness is  about 1060 counts, which is an increase of only 
about 70 counts compared to the bias level.

\section{Results}

As a first step, we identified a few stars in the field (see Fig.~1)
that can be used as comparison stars: \\
\begin{table*}
\caption{Coordinates and magnitudes of the comparison stars around LY~UMa.} 
\centering
\begin{tabular}{l l l  | l  l | c | c c c c}
 No. & &  object	    &  RA	 & Declination & $V$      &  $V$    &  $B$   & \\
     & &    		    &		 &	       & Tycho-2  &  GAIA   & GAIA   & \\ 
     & &    		    &		 &	       &          &         &        & \\
\# 1 & &  HD 233778	    & 10 51 44.4 & +52 05 23   & 9.79     &   9.853 &        & \\
\# 2 & &  TYC 3823-985-1    & 10 46 13.9 & +52 34 27   & 10.97    &  11.030 &        & \\
\# 3 & &  ...		    & 10 46 04.4 & +52 20 54   &          &  12.804 &        & \\ 
\# 4 & &  TYC 3449-829-1    & 10 48 29.7 & +52 26 48   & 10.86    &  10.879 & 11.500 & \\
\# 5 & &  ...		    & 10 48 39.7 & +52 13 17   &	  &  12.940 & 13.729 & \\
\# 6 & &  ...		    & 10 47 04.0 & +52 20 49   &          &  12.754 &        & \\ 
\# 7 & &  TYC 3449-702-1    & 10 47 30.5 & +52 11 20   & 12.48    &  12.337 & 12.849 & \\
\# 8 & &  ...               & 10 47 57.6 & +52 16 36   &          &  15.310 & 16.200 & \\
\# 9 & &  ...               & 10 47 51.8 & +52 15 43   &          &  14.876 & 15.532 & \\
\\
\label{t.cs}
\end{tabular}
\end{table*}

The $V$-band magnitudes are from Tycho-2 (H{\o}g et al. 2000) 
and GAIA (GAIA collaboration 2022, 2023) catalogues. 
The GAIA magnitudes are synthetic photometry taken from Vizier catalog I/360/syntphot. 
After a few measurements, we decided to use as comparison stars \#4, \#5, \#7, \#8,
adopting their  GAIA magnitudes.  
The measurements are presented in Table~1. 

\begin{table*}
\caption{$V$- and $B$-band photometry of LY~UMa with the 2.0m Rozhen telescope.} 
\centering
\begin{tabular}{ccc | c  c  cc | crc |  c cc cc c c c } 
Date   UT    & exp-time &        & Date   UT	& exp-time &        &  \\
             & [sec]    & $V$ mag & 		& [sec]    & $B$ mag &  \\
             &          &        & 		&	   &	    &  \\   
2024-04-13    23:37  &        60 &  15.248   & 2024-04-13  23:39  &  90   &   16.311  &  \\
2024-04-13    23:38  &        60 &  15.249   & 2024-04-13  23:40  &  90   &   16.301  &  \\
2024-04-13    23:39  &        60 &  15.244   & 2024-04-13  23:42  &  90   &   16.272  &  \\ 
2024-04-13    23:40  &        60 &  15.254   & 2024-04-13  23:43  &  90   &   16.261  &  \\
2024-04-13    23:41  &        60 &  15.247   & 2024-04-13  23:45  &  90   &   16.240  &  \\
2024-04-13    23:42  &        60 &  15.254   & 2024-04-13  23:46  &  90   &   16.240  &  \\ 
2024-04-13    23:43  &        60 &  15.243   & 2024-04-13  23:48  &  90   &   16.216  &  \\ 
2024-04-13    23:44  &        60 &  15.237   & 2024-04-13  23:49  &  90   &   16.195  &  \\
2024-04-13    23:45  &        60 &  15.245   & 2024-04-13  23:51  &  90   &   16.166  &  \\
2024-04-13    23:46  &        60 &  15.248   & 2024-04-13  23:52  &  90   &   16.229  &  \\
2024-04-13    23:47  &        60 &  15.246   & 2024-04-13  23:54  &  90   &   16.269  &  \\
2024-04-13    23:48  &        60 &  15.246   & 2024-04-13  23:56  &  90   &   16.269  &  \\
2024-04-13    23:49  &        60 &  15.232   & 2024-04-13  23:57  &  90   &   16.235  &  \\
2024-04-13    23:50  &        60 &  15.218   & 2024-04-13  23:59  &  90   &   16.300  &  \\
2024-04-13    23:51  &        60 &  15.221   & 2024-04-14  00:00  &  90   &   16.299  &  \\
2024-04-13    23:52  &        60 &  15.240   & 2024-04-14  00:02  &  90   &   16.252  &  \\
2024-04-13    23:53  &        60 &  15.252   & 2024-04-14  00:03  &  90   &   16.306  &  \\
2024-04-13    23:54  &        60 &  15.249   & 2024-04-14  00:05  &  90   &   16.320  &  \\
2024-04-13    23:55  &        60 &  15.256   & 2024-04-14  00:06  &  90   &   16.308  &  \\  
2024-04-13    23:56  &        60 &  15.260   & 2024-04-14  00:08  &  90   &   16.342  &  \\  
2024-04-13    23:57  &        60 &  15.251   & 2024-04-14  00:09  &  90   &   16.358  &  \\  
2024-04-13    23:58  &        60 &  15.268   & 2024-04-14  00:11  &  90   &   16.277  &  \\  
2024-04-13    23:59  &        60 &  15.281   &                    &  mean &  $16.271 \pm 0.047$ & \\  
2024-04-14    00:00  &        60 &  15.270   &  \\  
2024-04-14    00:01  &        60 &  15.280   &  \\  
2024-04-14    00:02  &        60 &  15.270   &  \\  
2024-04-14    00:03  &        60 &  15.288   &  \\  
2024-04-14    00:04  &        60 &  15.300   &  \\  
2024-04-14    00:06  &        60 &  15.289   &  \\  
2024-04-14    00:07  &        60 &  15.294   &  \\  
2024-04-14    00:08  &        60 &  15.307   &  \\  
2024-04-14    00:09  &        60 &  15.303   &  \\  
2024-04-14    00:10  &        60 &  15.303   &  \\  
2024-04-14    00:11  &        60 &  15.312   &  \\  
2024-04-14    00:12  &        60 &  15.311   &  \\  
                     &    mean   &  $15.263 \pm 0.031$ &  \\ 
\end{tabular}					        					      
\end{table*}

\begin{table*}
\caption{$V$ band photometry of LY~UMa and Mrk~153 with the 50/70 cm Rozhen (2025-04-28) 
and 60~cm Shumen telescopes (2025-06-02).} 
\centering
\begin{tabular}{ccc | c  c  cc | crc |  c cc cc} 
telescope & Date   UT     & exp-time & LY~UMa & Mrk~153 & \\
              & [sec]    &        &         & \\
              &          &        &         & \\   
50/70~cm $^{} \;$ & 2025-04-28  21:45 &  60 & 15.413           & 14.847 & \\
         & 2025-04-28  21:48 &  60 & 15.438           & 14.813 & \\
         & 2025-04-28  21:51 &  60 & 15.420           & 14.776 & \\
         & 2025-04-28  21:55 &  60 & 15.402           & 14.821 & \\
         & 2025-04-28  21:58 &  60 & 15.461           & 14.811 & \\  
         &   & mean            & $15.427\pm0.023$ & $\; 14.81 \pm 0.03$ & \\  
                  &     &                  &        & \\  
60~cm  & 2025-06-02 20:04  & 120 & 15.588           &        & \\
       & 2025-06-02 20:06  & 120 & 15.556           &        & \\
       & 2025-06-02 20:08  & 120 & 15.556           &        & \\
       &   & mean           & $15.567\pm0.008$ &        & \\ 
\\
\end{tabular}
\caption{$B$ band photometry of LY~UMa with the 50/70~cm Rozhen telescope. }   
\begin{tabular}{ccc | c  c  cc | crc |  c cc cc} 
Date   UT     & exp-time      &       LY~UMa         & \\
              & [sec]         &      $B$ mag           & \\
2025-04-28 22:28:03  &  90.0  &  $16.559 \pm 0.018$  & \\
2025-04-28 22:31:02  &  90.0  &  $16.674 \pm 0.021$  & \\
2025-04-28 22:34:01  &  90.0  &  $16.553 \pm 0.019$  & \\
2025-04-28 22:37:00  &  90.0  &  $16.554 \pm 0.020$  & \\
2025-04-28 22:39:59  &  90.0  &  $16.560 \pm 0.026$  & \\
2025-04-28 22:43:11  &  90.0  &  $16.446 \pm 0.019$  & \\
2025-04-28 22:46:10  &  90.0  &  $16.596 \pm 0.005$  & \\
2025-04-28 22:49:09  &  90.0  &  $16.606 \pm 0.015$  & \\
2025-04-28 22:52:08  &  90.0  &  $16.504 \pm 0.018$  & \\
2025-04-28 22:55:07  &  90.0  &  $16.601 \pm 0.011$  & \\
                     &  mean  &  $16.565 \pm 0.061$  & \\
\end{tabular}
\end{table*}

\subsection{LY~UMa}

For LY~UMa SIMBAD gives $V= 14.95$, GAIA  gives   $V=15.28$.
Kazarovets et al. (2006) reported  that the object varies in the range 
$15.44 > V > 14.95$. Our observations give range: \\
$15.31 > V > 15.22$ for our run on 13/14 April 2024, \\
$15.46 > V > 15.40$ for our run on 28 April 2025, \\ 
$15.59 > V > 15.56$ for our run on 2 June  2025. \\
In our runs visual brightness of LY~UMa is in the range $15.59 > V > 15. 22$. 
Its $B$ band magnitude is in the range $16.67 > B > 16.17$.  
On 13/14 April 2024, we estimate average $B=16.27$, $V=15.26$, and colour $B-V=1.01$.
On 28 April 2025, we estimate average $B=16.57$, $V=15.43$, and colour $B-V=1.14$.
As expected when the star is brighter it is  bluer.

LY~UMa is at a distance $d=310 \pm 2$~pc (Bailer-Jones et al. 2021).
There is no interstellar extinction toward the object: \\
$^{} \; \; \; ^{}$ https://astro.acri-st.fr/gaia\_dev/\#extinction (Lallement et al.\ 2019). \\
Using  the standard formula
\begin{equation}
M = m -  5 \, log_{10}(d [pc]) + 5, 
\end{equation}
we estimate the absolute $V$-band magnitude of LY~UMa during our observations $M_V =7.94-8.10$. 
The spectral observations of LY~UMa 
demonstrate that it is an emission-line binary with a strong late-K contribution (Tappert et al. 2001).
Therefore,  we suppose that the K star contributes $70\% - 95 \%$ in the $V$-band.
In this way, we estimate the absolute $V$-band magnitude of the K star $M_V =8.49 - 8.00$.

The  "Modern Mean Dwarf Stellar Color and Effective Temperature Sequence"
by  Pecaut \& Mamajek (2013) [\footnote{we use  version 2022.04.16 as given at\\
www.pas.rochester.edu/$\sim$emamajek/EEM\_dwarf\_UBVIJHK\_colors\_Teff.txt}]  
gives the following absolute $V$-band magnitudes for K-type main-sequence stars:  
for K6V  $M_V =7.64$, for K7V  $M_V =8.16$, for K8V  $M_V =8.43$, for K9V  $M_V =8.56$. 
Our photometric observations indicate 
that most likely the mass donor of LY~UMa is of spectral type K7V. 
Following  Pecaut \& Mamajek (2013), 
a typical K7V star has 
$M_V=8.16$, colour $B-V = 1.34$,  radius $R = 0.63 \; R_\odot$, and mass $M=0.64 \: M_\odot$.
As expected the $B-V$ index of LY~UMa ($B-V= 1.01 - 1.14$) is bluer, 
due to the contribution of the accretion flow to the optical brightness.

\subsection{Mrk~153}

The Active Galaxy Nucleus Candidate Mrk 153 (Petrosian et al. 2007) 
is located at 7 arcmin  from our primary target LY~UMa.
Such objects are potentially variable, and we provide our estimate
of its brightness $V = 14.81 \pm 0.03$ on 2025-04-28T21:51.
The nucleus of the galaxy is $\approx 0.11$ magnitudes fainter.  \\

\section{Discussion}

Tappert et al. (2001) discussed several scenarios for the nature  of LY~UMa: 

1. nova-like - the nova-like CVs are in a permanent state of high mass transfer;

2. dwarf nova - the dwarf novae are in lower mass transfer state and
demonstrate outbursts due to disc instability; 

3. magnetic CV - in which the white dwarf is highly magnetized. 

Tappert et al. (2001) found that a dwarf nova classification is 
the most probable, although it does not completely explain the observed characteristics. 
If LY~UMa is really a dwarf nova, it probably  has 
a long period of accumulation of material in the accretion disc. 
Our results should be useful for future studies
of the binary nature of  LY~UMa.

\vskip 0.2cm 

\section{ Conclusions }
We performed $B$- and $V$-band photometry 
with the 50/70~cm Rozhen, 2.0m Rozhen and 60~cm Shumen telescopes 
of the cataclysmic variable LY~UMa, and find that its brightness is in the range
$V = 15.59 - 15.23$ 
and colour  $B-V = 1.14 - 1.01$. 
Using the GAIA distance 
we estimate absolute $V$-band magnitude of LY~UMa $7.94 \le  M_V \le 8.10$. 
For the mass donor we find spectral type K7-K8V
and absolute $V$-band magnitude  $8.0 \le  M_V \le 8.5$.
For the Active Galaxy Nucleus Mrk~153 we observed apparent magnitude 
$V = 14.81 \pm 0.03$.

\vskip 0.3cm 

{\bf Acknowledgments: }
We devote this paper to the memory of our dear colleague Lyuba Dankova (1971-2026).
She was a very keen and dedicated astronomer who performed observations 
and led this scientific work.  
This work is part of the project KP-06-H98/8 "Accretion flows in binary stars"
(Bulgarian National Science Fund).  
We acknowledge the National Roadmap for Scientific Infrastructure 
coordinated by Ministry of Education and Science of Bulgaria. 
DM acknowledges support from Science Fund of Shumen University. 
We thank Dr. Petia Yanchulova Merica-Jones for a thorough reading of the manuscript. 
We thank an anonymous referee for making very valuable suggestions.

\end{document}